\documentclass[10pt,twocolumn,letterpaper]{article}

\usepackage[applications]{wacv}      
\usepackage{multirow}
\usepackage{booktabs} 
\usepackage{xcolor}
\usepackage{colortbl}
\newcommand{\shline}{\specialrule{1pt}{0pt}{0pt}} 
\definecolor{wacvblue}{rgb}{0.21,0.49,0.74}
\usepackage[pagebackref,breaklinks,colorlinks,allcolors=wacvblue]{hyperref}

\def\wacvPaperID{465} 
\def\confName{WACV}
\def\confYear{2026}

\title{Multimodal Medical Image Binding via Shared Text Embeddings}

\author{Yunhao Liu\textsuperscript{*}\\
The Hong Kong Polytechnic University\\
Hong Kong, China\\
{\tt\small yunhao.liu@connect.polyu.hk}
\and
Suyang Xi\textsuperscript{*}\\
Emory University\\
Atlanta, USA\\
{\tt\small }
\and
Shiqi Liu\textsuperscript{*}\\
The University of Hong Kong\\
Hong Kong, China\\
{\tt\small }
\and
Hong Ding\\
University of Illinois Chicago\\
Chicago, USA\\
{\tt\small }
\and
Chicheng Jin\\
University of Science and Technology of China\\
Hefei, China\\
{\tt\small }
\and
Chong Zhong\\
The Hong Kong Polytechnic University\\
Hong Kong, China\\
{\tt\small }
\and
Junjun He\\
Shanghai AI Laboratory\\
Shanghai, China\\
{\tt\small }
\and
Catherine C. Liu\textsuperscript{$\dagger$}\\
The Hong Kong Polytechnic University\\
Hong Kong, China\\
{\tt\small catherine.chunling.liu@connect.polyu.hk}
\and
Yiqing Shen\textsuperscript{$\dagger$}\\
Johns Hopkins University\\
Baltimore, USA\\
{\tt\small yiqingshen1@gmail.com}
\thanks{Equal contribution.}
\thanks{Corresponding author.}
}

\begin{document}
\maketitle
\begin{abstract}
Medical image analysis increasingly relies on the integration of multiple imaging modalities to capture complementary anatomical and functional information, enabling more accurate diagnosis and treatment planning.
Achieving aligned feature representations across these diverse modalities is therefore important for effective multimodal analysis.
While contrastive language-image pre-training (CLIP) and its variant have enabled image-text alignments, they require explicitly paired data between arbitrary two modalities, which is difficult to acquire in medical contexts. 
To address the gap, we present Multimodal Medical Image Binding with Text (M\textsuperscript{3}Bind), a novel pre-training framework that enables seamless alignment of multiple medical imaging modalities through a shared text representation space without requiring explicit paired data between any two medical image modalities.
Specifically, based on the insight that different images can naturally bind with text, M\textsuperscript{3}Bind first fine-tunes pre-trained CLIP-like image-text models, which are derived from different medical modalities, to align their modality-specific text embedding space while preserving their original image-text alignments. 
Subsequently, we distill these modality-specific text encoders into a unified model, creating a shared text embedding space.
Notably, M\textsuperscript{3}Bind is a flexible framework in which the selection of CLIP-like models is not fixed and can be adapted according to the requirements of the task.
Experiments on X-ray, CT, retina, ECG, and pathological images on multiple downstream tasks demonstrate that M\textsuperscript{3}Bind achieves competitive or even superior performance in zero-shot, few-shot classification and cross-modal retrieval tasks compared to its CLIP-like counterparts.
These results validate M\textsuperscript{3}Bind's effectiveness in achieving cross-image-modal alignment for medical analysis.
%
\end{abstract}
    
\section{Introduction}
\label{sec:intro}
\begin{figure}[h]
    \centering
    \includegraphics[width=0.5\textwidth]{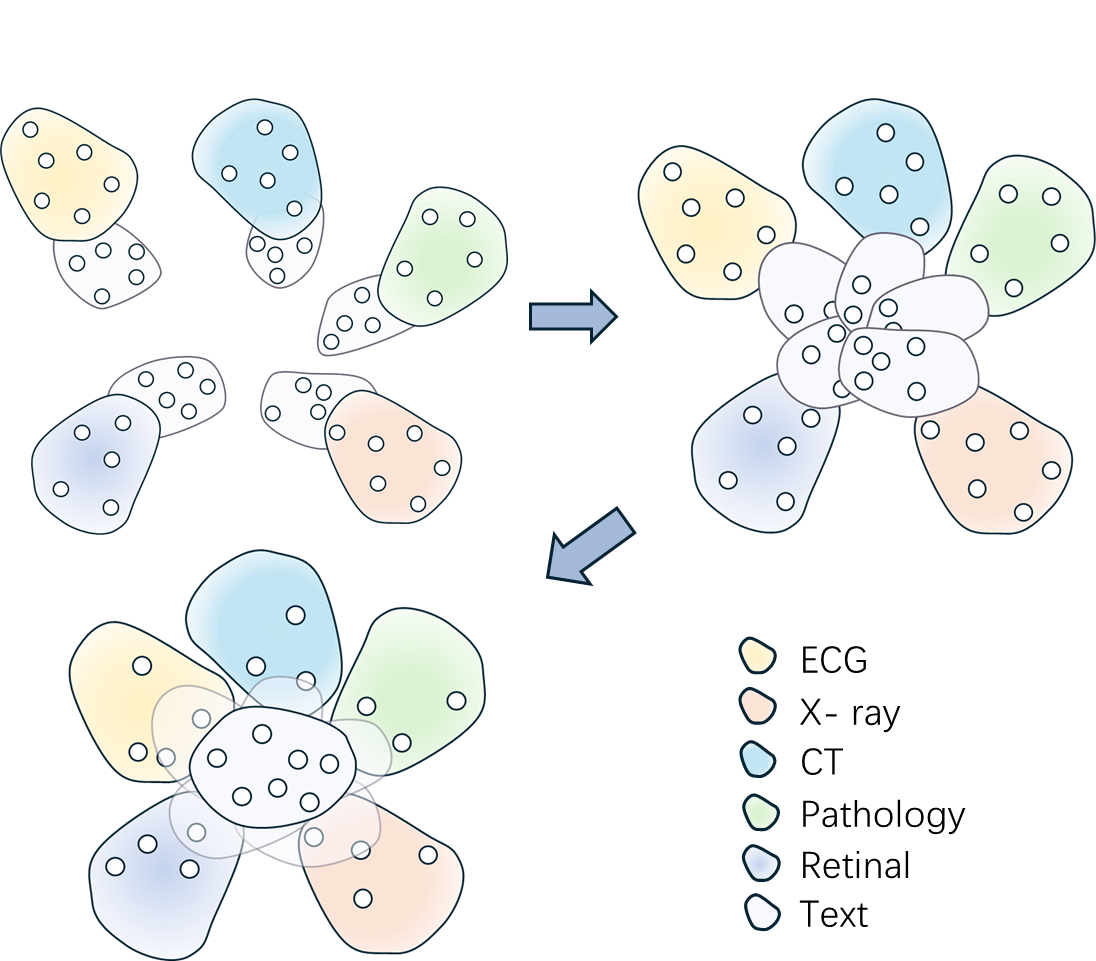} 
    \caption{We illustrate the M$^3$Bind framework, designed to embed multiple medical imaging modalities (X-ray, CT, retina, ECG, and Pathology) within a cohesive text-anchored space. In this figure, the transition from Phase One to Phase Two represents the five modality-specific CLIP-like models we used. These models start by aligning each modality with the text modality in isolation and then use text as the anchor modality, achieving joint alignment of all modalities through alignment with different text encoders. The progression from Phase Two to Phase Three indicates the further distillation of knowledge from the five aligned modality-specific text encoders into a unified text encoder, to achieve a consistent representation of all texts paired with different modalities. }
    \label{fig:intro_image}
\end{figure}

Medical diagnosis is increasingly based on the integration of multiple imaging modalities that provide complementary anatomical and pathological information \cite{hussain2022modern}. 
For example, computed tomography (CT) enables high resolution 3D tissue visualization \cite{rubin2014computed}, chest radiographs offer a broad structural context \cite{speets2006chest}, retina imaging captures microvascular patterns \cite{liew2008retinal}, electrocardiogram (ECG) records the electrical activity of the heart \cite{mirvis2001electrocardiography}, and histopathology provides tissue visualization at the cell level \cite{majno2004cells}. 
While each modality contributes unique diagnostic information, their full potential emerges when analyzed together in a multimodal fashion \cite{davnall2012assessment}.
For example, the evaluation of respiratory diseases benefits from integrating the detailed visualization of the lung parenchyma of CT with the broader anatomical context provided by X-rays \cite{chicklore2013quantifying}.
Similarly, cardiovascular diagnostics leverage the synergy between retinal vasculature patterns and histopathological tissue examination to evaluate both structural and functional aspects of cardiac health \cite{o2017imaging}. 
Conditions such as pericarditis require the combination of ECG data showing changes in the ST segment and PR segment with chest radiographs that reveal the cardiac silhouette or pericardial calcification \cite{10.1093/eurheartj/ehv318}.
These clinical scenarios highlight the importance of aligned feature representations that can effectively synthesize diagnostic information across diverse medical imaging modalities \cite{glynne2014anal}.

Contrastive language image pre-training (CLIP) \cite{radford2021learning} can align embedding spaces between visual and textual information through contrastive learning \cite{li2022blip, li2022grounded, mu2022slip, yao2021filip}. 
In medical image analysis, specialized CLIP variants have been proposed to integrate specific medical images with the corresponding clinical text descriptions \cite{zhao2023clip}. 
For radiology applications, GLoRIA \cite{huang2021gloria} and MGCA \cite{wang2022multi} improve diagnostic interpretability by aligning chest X-rays with radiology reports, while CT-CLIP \cite{hamamci2024foundation} extends this framework to volumetric CT data analysis.
In ophthalmology and pathology domains, FLAIR \cite{silva2025foundation} and QuiltNet \cite{ikezogwo2023quilt1m} respectively leverage semantic supervision for zero-shot retinal diagnosis and fine-grained histological interpretation.
However, these approaches typically focus on alignment between a single medical imaging modality and text, limiting their applicability to multiple medical imaging modalities \cite{wang2022multi, huang2021gloria}. 
For $n$ different modalities, conventional CLIP-based approaches would require training ${n \choose 2} = \frac{n(n-1)}{2}$ separate alignment models, which is not only computationally intensive but also limited by the difficulty of acquiring the corresponding paired data between different imaging types.

Recent work in multimodal learning demonstrates that diverse modalities can be aligned in a shared embedding space without requiring explicit pairwise training. 
Specifically, ImageBind \cite{girdhar2023imagebind} uses natural images as a bridge to connect six different data modalities through their inherent connections to visual information. 
However, direct application of ImageBind to medical imaging is not applicable due to the lack of an anchor visual modality.
Alternatively, ``\textit{binding through images}'' concept inspires our approach to multimodal medical imaging, where diverse modalities frequently co-occur with clinical text descriptions that can serve as semantic bridges.
For example, radiological images are routinely accompanied by detailed reports, pathology slides with diagnostic annotations, and retina scans with clinical evaluations. This motivates the use of clinical text as a semantic anchor to align diverse medical imaging modalities in a unified representation space (Figure~\ref{fig:intro_image}).

To bridge the gap, we propose multimodal medical image binding with text (M\textsuperscript{3}Bind), a multimodal pre-training framework for medical image analysis. 
M\textsuperscript{3}Bind innovates by its fine-tuning of CLIP-like models across multiple modalities, using text as an anchor modality to bind to various medical imaging modalities. 
Unlike previous CLIP-like methods, M\textsuperscript{3}Bind creates a shared text representation space by aligning the text modalities of all CLIP-like models, thereby aligning different image modalities without requiring explicit paired data between any two image modalities. 
We further employ distillation \cite{hinton2015distilling, jiao2019tinybert, mirzadeh2020improved, sanh2019distilbert, touvron2021training, wu2022tinyvit} to consolidate modality-specific text encoders into a unified model that enables seamless cross-modal analysis.

The main contributions are four-fold.
Firstly, we propose M\textsuperscript{3}Bind, a novel multimodal pre-training framework that addresses the challenge of cross-modal medical image alignment by using text as a central bridging modality, enabling seamless and flexible integration of multiple CLIP-like image-text encoders without requiring explicit paired data between image modalities.
Secondly, we propose an adaptive training strategy that effectively counters the inherent modality imbalance between different medical image modalities, ensuring balanced representation learning across diverse modalities.
Thirdly, we introduce Shared Embedding Space Knowledge Distillation to consolidate representations from multiple modality-specific text encoders into a single unified text model, enhancing computational efficiency while preserving alignment quality across modalities.
Finally, we conduct a comprehensive evaluation of M\textsuperscript{3}Bind on 12 diverse datasets that span X-ray, CT, retina, ECG, and pathological images, demonstrating state-of-the-art performance in zero shot classification, few shot classification, and cross-modal retrieval tasks compared to existing CLIP-based approaches.

\section{Related Work}
\label{sec:Relatedwork}
\subsection{CLIP Variants} 

CLIP has inspired numerous adaptations that improve multimodal representation learning in diverse domains. 
Early attempts focused on representation quality, where SLIP \cite{mu2022slip} incorporated self-supervised learning objectives alongside contrastive training to improve feature robustness, while DeCLIP \cite{smeu2024declip} leveraged multi-dimensional supervision signals to increase data efficiency.
Several later approaches have explored masking strategies to optimize the alignment process.
For example, MaskCLIP \cite{dong2023maskclip} employed masked self-distillation with an exponential moving average (EMA) encoder to strengthen alignment quality and noise resilience, while FLIP \cite{li2023scaling} and Alpha-CLIP \cite{sun2024alpha} introduced input masking that balances computational efficiency with model depth and performance. 
Most relevant to our work, ImageBind \cite{girdhar2023imagebind} advanced multimodal integration by unifying six distinct modalities, including images, text, audio, depth, thermal, and IMU data, into an aligned representation space using images as a bridge modality.
Although ImageBind demonstrated impressive performance in tasks in the natural domain, its approach cannot be directly applied to medical imaging contexts where no single image modality can serve as a universal anchor.

\subsection{Medical CLIP-like Models}
Medical imaging has seen adaptation of CLIP to address domain-specific challenges, which leverage contrastive learning to align medical images with clinical text.
For volumetric data analysis, CT-CLIP \cite{hamamci2024foundation} aligns 3D computed tomography scans with radiology reports.
%
%
Ophthalmology applications benefit from FLAIR \cite{silva2025foundation}, which enriches the analysis of retinal images through the supervision of semantic text, substantially improving zero shot diagnostic capabilities. 
For chest radiography, MGCA \cite{wang2022multi} proposes multi-granularity cross-modal contrastive learning to capture hierarchical disease manifestations. 
In digital pathology, QuiltNet \cite{ikezogwo2023quilt1m} integrates histopathological images with expert knowledge for the interpretation of fine-grained tissue.
Despite these advances, each model operates within a single imaging modality, limiting their utility in comprehensive multimodal analysis scenarios. 
BiomedCLIP\cite{zhang2023biomedclip} and UniMed-CLIP\cite{khattak2024unimed} aim to collect more diverse and extensive data to train models. However, these methods employ a single shared visual encoder for images across different modalities. Given the significant heterogeneity among multimodal medical images, this approach inevitably limits the representational capacity of the model. 
MedBind \cite{gao2024medbind} represents an initial step toward multimodal integration in medicine by introducing a specialized loss function to handle paired X-ray and ECG data.
However, MedBind still relies on explicitly curated image-to-image pairings, which are difficult to obtain at scale across diverse medical imaging modalities. 
This dependency on paired data presents a constraint for broader multimodal medical image analysis.
Our proposed M\textsuperscript{3}Bind addresses this limitation by introducing a text-centric binding approach that eliminates the need for direct image-to-image pairings.

\begin{figure*}[t]
    \centering
    \includegraphics[width=0.95\linewidth]{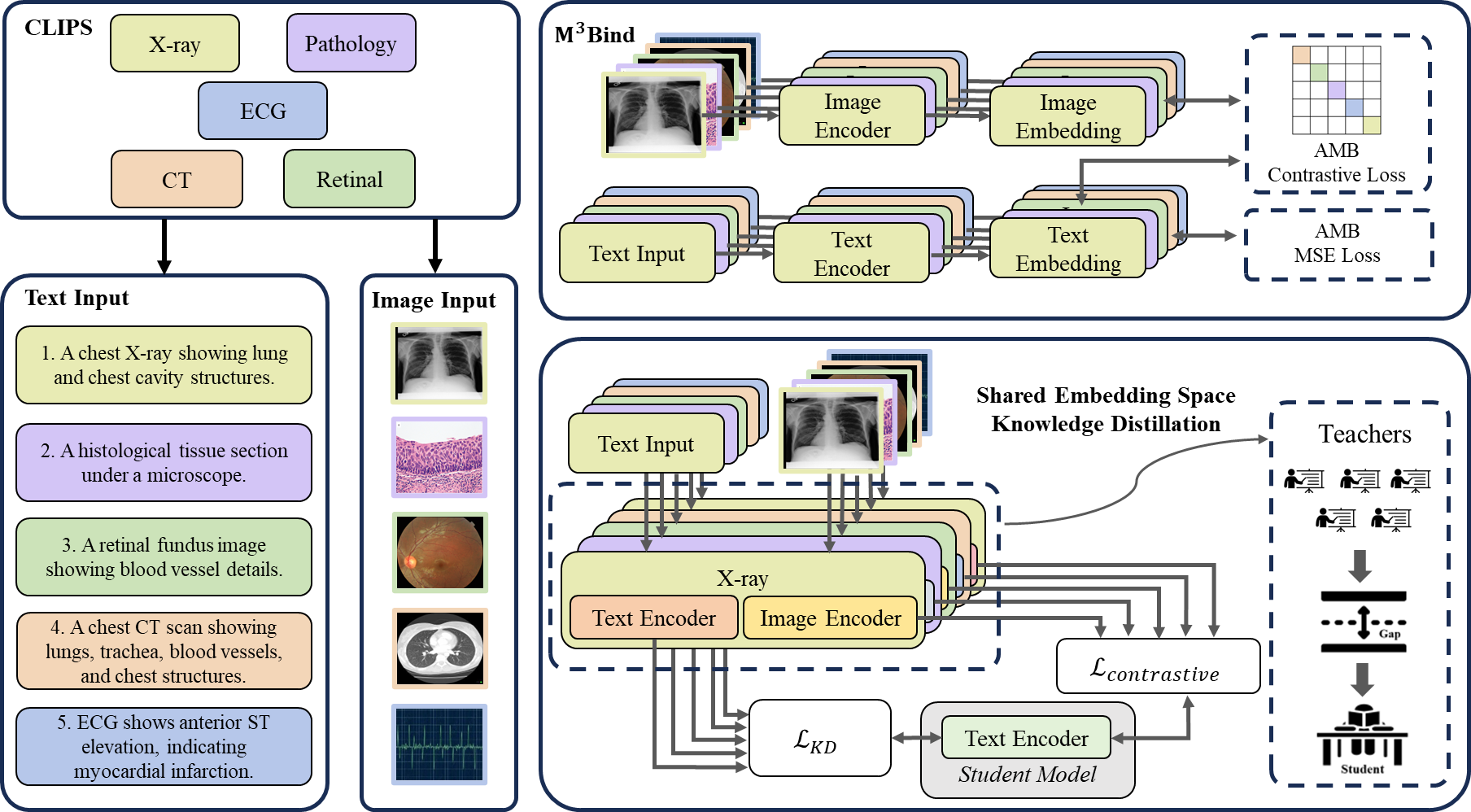} 
    \caption{Overview of the M\textsuperscript{3}Bind. 
    We start with five modality-specific CLIP-like models (X-ray, CT, retina, ECG, and pathology). 
    The M\textsuperscript{3}Bind integrates these models by maintaining image-text alignment within each modality while aligning text representations across different modalities. 
    LoRA is applied to efficiently fine-tune these pre-trained models while preserving their original capabilities. 
    The Shared Embedding Space Knowledge Distillation (SESKD) consolidates knowledge from all modality-specific text encoders (teachers) into a unified BioClinicalBERT-based student model. 
    This student model is trained using both MSE loss for direct embedding alignment and contrastive loss to preserve semantic relationships with all five imaging modalities.
}
    \label{fig:method}
\end{figure*}

\subsection{Knowledge Distillation}
In single-modality contexts, Knowledge Distillation (KD) achieved notable success by compressing large networks while preserving performance through the transfer of soft probability distributions rather than hard labels\cite{hinton2015distilling}. 
Regarding multimodal fields, recent attempts include DistillVLM \cite{wang2023efficient}, which introduced cross-modal attention alignment to preserve vision-language interactions between teacher and student models.
Additionally, TinyCLIP \cite{wu2023tinyclip} used affinity mimicking and weight inheritance to optimize vision language embeddings, although its effectiveness remains constrained by architectural similarity requirements. 
CLIP-KD \cite{yang2024clip} overcomes this limitation by supporting knowledge transfer between heterogeneous architectures.
Further advances have refined the granularity of knowledge transfer in multimodal models. 
DIME-FM \cite{sun2023dime} proposed instance-level distillation to capture fine-grained relationships between individual samples, while CLIP-CID \cite{yang2024clipcid} extended this approach with cluster-level distillation that preserves semantic relationships between instance groups.
In our M\textsuperscript{3}Bind, we leverage KD to address the challenge of unifying multiple modality-specific text encoders into a single model while preserving alignment with diverse medical imaging modalities.

\section{Methods}

\subsection{Method Overview}
M\textsuperscript{3}Bind aims to achieve the cross-modal medical image alignment through a novel text-bridged architecture that eliminates the need for explicitly paired data between imaging modalities. 
As illustrated in Fig.~\ref{fig:method}, M\textsuperscript{3}Bind consists of three components working in concert to achieve seamless multimodal integration. 
First, we leverage specialized CLIP-like models pre-trained on individual medical imaging modalities (X-ray, CT, retina, ECG, and pathology) and align their text embedding spaces while preserving their original image-text relationships. 
This alignment enables any two medical imaging modalities to be semantically connected through their shared text representation without requiring direct cross-modal image pairs, which are typically difficult to acquire in medical contexts. 
Second, we propose an adaptive modality balancing strategy that counteracts the inherent data imbalance across different medical imaging domains through dynamic sampling, modality-specific learning rate scaling, and weighted loss functions. 
This ensures that smaller datasets receive appropriate emphasis during training, preventing dominance by more abundant modalities such as X-rays. 
Third, we employ Shared Embedding Space Knowledge Distillation (SESKD) to consolidate knowledge from multiple modality-specific text encoders into a single unified BioClinicalBERT-based model, reducing computational overhead while maintaining cross-modal alignment quality. 
Together, these components enable M\textsuperscript{3}Bind to create a unified semantic space where diverse medical imaging modalities can be effectively integrated for comprehensive clinical analysis and interpretation.

\subsection{M\textsuperscript{3}Bind for Pre-training}

M\textsuperscript{3}Bind introduces a novel pertaining method that leverages text as a universal anchor to align five distinct medical imaging modalities (X-ray, CT, retina, ECG, and pathology) without requiring paired image data between image modalities.
Specifically, M\textsuperscript{3}Bind starts with CLIP-like medical models, each pre-trained on a specific medical imaging modality (X-ray, CT, retina, ECG, and pathology) with corresponding textual descriptions. 
Building upon five CLIP variants namely MGCA \cite{wang2022multi}, CT-CLIP \cite{hamamci2024foundation}, FLAIR \cite{silva2025foundation}, ECG-CLIP (proposed by MedBind\cite{gao2024medbind} and trained by ourselves), and QuiltNet \cite{ikezogwo2023quilt1m},  M\textsuperscript{3}Bind begins with these modality-specific pre-trained image-text models, each trained on their respective medical imaging modality with corresponding textual descriptions. The selection of these models is flexible and can be freely chosen according to specific needs, not limited to the models we have actually selected.
The first stage of M\textsuperscript{3}Bind establishes text as a shared semantic space by aligning the text embedding spaces of these five CLIP-like models while preserving their respective original image-text alignments through contrastive learning. 
Specifically, for each modality $m$, we maintain the alignment of image and text using the following contrastive loss:
\begin{equation} 
\mathcal{L}_{CLIP}^{(m)} = - \sum_{i=1}^{N} 
\log \frac{\exp \left( \text{sim} \left( \mathbf{I}_i^{(m)}, \mathbf{T}_i^{(m)} \right) / \tau \right)}{\sum_{j=1}^{N} \exp \left( \text{sim} \left( \mathbf{I}_i^{(m)}, \mathbf{T}_j^{(m)} \right) / \tau \right)},\label{eq:cliploss} 
\end{equation} 
where $\mathbf{I}_i^{(m)}$ represents the image embedding of modality $m$ and $\mathbf{T}_i^{(m)}$ denotes the corresponding text embeddings, $\tau$ is the temperature scaling factor, and $N$ denotes the batch size of paired image-text samples.
To align the text embedding spaces across different modalities, we employ a mean squared error (MSE) loss between arbitrary modality-specific text encoders, namely
\begin{equation}
\mathcal{L}_{MSE}^{(m_1,m_2)} = 
\frac{1}{N} \sum_{i=1}^{N} \left| \mathbf{T}_{m_1,i} - \mathbf{T}_{m_2,i} \right|^2 
\end{equation} 
where $\mathbf{T}_{m_1,i}$ and $\mathbf{T}_{m_2,i}$ represent text embeddings with respect to different image modalities $m_1$ and $m_2$ corresponding to the same semantic content $i$. 
By minimizing this distance, we encourage the convergence of diverse modality-specific text representations from different text encoders into an aligned space.

\subsection{Adaptive Modality Balancing}
Medical datasets with respect to different modalities vary substantially in size, with X-ray collections often containing millions of images, while ECG and retinal images may be orders of magnitude smaller. 
To address this inherent modality imbalance, M\textsuperscript{3}Bind incorporates an adaptive modality balancing strategy with three components.
First, we employ an adaptive sampling approach that dynamically adjusts the probability of selecting text data from each modality during training.  
For any modality $m$, the sampling probability is inversely proportional to its dataset size $|D_m|$, namely \begin{equation}
p_m = \frac{\left(1/|D_m|\right)^\beta}{\sum_{m=1}^{M} \left(1/|D_m|\right)^\beta}, \end{equation} 
where $\beta$ is the balancing hyperparameter that controls the degree of compensation for underrepresented modalities. 
Higher values of $\beta$ increase the sampling frequency of smaller datasets, while $\beta=0$ would result in uniform sampling regardless of dataset size. We empirically set $\beta=0.5$ in our experiments.
Second, we implement modality-specific learning rate scaling to optimize convergence across datasets of varying sizes. 
The learning rate $\eta_m$ for each modality $m$ is adjusted according to
\begin{equation} 
\eta_m = \frac{\eta_0}{\sqrt{|D_m|}}, \end{equation} 
where $\eta_0$ represents the base learning rate. 
It allows smaller datasets to benefit from more aggressive parameter updates while preventing overfitting on larger datasets through more conservative updates.
Third, we incorporate modality-specific loss weighting to ensure balanced optimization across all imaging types. 
Each modality is assigned a weight inversely proportional to its dataset size, namely 
\begin{equation} 
w_m = \frac{1}{\sqrt{|D_m|}} .
\end{equation} 
These weights are applied to the total objective function, ensuring that smaller datasets receive appropriate emphasis during training.
The complete training objective integrates these balancing mechanisms into: 
\begin{equation} 
\mathcal{L}_{total} = \sum_{m=1}^{M} w_m \cdot \mathcal{L}_{CLIP}^{(m)} + \lambda \cdot w_{m_1} \cdot w_{m_2} \cdot \sum_{m_1 \neq m_2} \mathcal{L}_{MSE}^{(m_1,m_2)}. \label{eq:loss}
\end{equation}
where the first term represents the weighted sum of modality-specific CLIP losses in Eq.~\eqref{eq:cliploss}, maintaining intra-modality alignment, and the second term aggregates MSE losses between all pairs of modalities $(m_1, m_2)$ to enforce alignment between the text embeddings. 
The hyperparameter $\lambda$ is the balancing coefficient.
To preserve the well-established representational capabilities of the pre-trained CLIP-like models while efficiently adapting them to our multi-modal context, we leverage Low-Rank Adaptation (LoRA) \cite{hu2021lora} in optimizing Eq.~\eqref{eq:loss}.

\subsection{Shared Embedding Space Knowledge Distillation}

To consolidate the text embedding space into a single unified model, we propose Shared Embedding Space Knowledge Distillation (SESKD).
The SESKD employs modality-specific text encoders as teachers, including ones for X-ray, Retina, Pathology, ECG, and CT, and distills their collective knowledge into a single student model based on BioClinicalBERT \cite{alsentzer2019publicly}. 
We choose BioClinicalBERT as it has been pre-trained specifically on biomedical and clinical text, providing domain-appropriate linguistic representations that align with the medical context of our imaging modalities.
Our distillation process operates through two objectives. 
First, we apply a MSE loss to align the student's text embeddings with each teacher encoder's output, namely
\begin{equation}
\mathcal{L}_{\text{KD}} = \sum_{m=1}^{M} \text{MSE}(\mathbf{T}_m, \mathbf{T}_{\text{student}}),
\end{equation}
where $\mathbf{T}_m$ represents each teacher text encoder's output embedding, and $\mathbf{T}_{\text{student}}$ is the corresponding embedding from the student model. 
However, simply matching embeddings is insufficient to preserve the alignment between text and image modalities. 
Therefore, we incorporate a contrastive learning objective that enforces semantic coherence between the student's text embeddings and the image embeddings from each modality's image encoder, namely
\begin{equation}
\mathcal{L}_{\text{Contrastive}} = -\sum_{i=1}^{N} \log \frac{\exp(\text{sim}(\mathbf{T}_{\text{student}}, \mathbf{I}_{i}) / \tau)}{\sum_{j=1}^{N} \exp(\text{sim}(\mathbf{T}_{\text{student}}, \mathbf{I}_{j}) / \tau)},
\end{equation}
where $\text{sim}(\mathbf{T}_{\text{student}}, \mathbf{I}_{i})$ denotes the cosine similarity between the student's text embedding $\mathbf{T}_{\text{student}}$ and the corresponding image embedding $\mathbf{I}_{i}$ from the respective teacher's image encoder, with $\tau$ serving as a temperature parameter to control the sharpness of the similarity distribution.
The complete distillation objective integrates both components, depicted as
\begin{align}
\mathcal{L}_{\text{SESKD}} = \mathcal{L}_{\text{KD}} + \mathcal{L}_{\text{Contrastive}},
\end{align}

\begin{table*}[h]
\centering
\setlength{\abovecaptionskip}{0pt} 
\setlength{\belowcaptionskip}{0pt} 
\caption{Zero-Shot Performance Comparison Across Four Modalities (Retina, X-Ray, CT, Pathology) Using M$^3$Bind and Baseline Models.}
\label{tab:sys_compare}

\resizebox{\linewidth}{!}{
\begin{large}
\begin{tabular}{l|cccc|c|ccc|c|cccc}
\shline
\multirow{3}{*}{\textbf{Methods}} & \multicolumn{4}{c|}{\textbf{Retina}} & \multirow{3}{*}{\textbf{Methods}} & \multicolumn{3}{c|}{\textbf{X-Ray}} & \multirow{3}{*}{\textbf{Methods}}& \multicolumn{4}{c}{\textbf{Retina}}\\ \cline{2-5} \cline{7-9}\cline{11-14}
 & \multicolumn{2}{c|}{\textbf{MESIDOR}\cite{decenciere2014feedback}} & \multicolumn{2}{c|}{\textbf{ODIR200x3}\footref{1}} & & \multicolumn{3}{c|}{\textbf{CheXpert 200x5}\cite{irvin2019chexpert}}  && \multicolumn{2}{c|}{\textbf{MIMIC-ECG}\cite{gow2023mimic}} & \multicolumn{2}{c}{\textbf{PTB-XL}\cite{wagner2020ptb}}\\ \cline{2-5} \cline{7-9}\cline{11-14}
 & \textbf{$\pi_{\text{naive}}$} & \textbf{$\pi_{EK}$} & \textbf{$\pi_{\text{naive}}$} & $\pi_{EK}$ & & \textbf{Accuracy} & \textbf{F1-Score} & \textbf{Precision}&& \textbf{$R_1$} & \textbf{$R_{10}$} & \textbf{$R_1$} & \textbf{$R_{10}$}\\

\shline
CLIP\cite{radford2021learning} & 0.237 & 0.200 & 0.445 & 0.480 & ConVIRT\cite{zhang2022contrastive} & 21.30 & 19.03 & 17.65 & -- & -- & -- & -- & -- \\
BiomedCLIP\cite{zhang2023biomedclip} & 0.224 & 0.207 & \textcolor{red}{0.727} & 0.583 & GLORIA\cite{huang2021gloria} & 23.20 & 15.79 & \textcolor{red}{44.01}& ECG-CLIP & 51.7 & \textcolor{blue}{95.5} & \textcolor{blue}{2.2} & 17.5 \\
FLAIR-$\pi_{\text{naive}}$\cite{silva2025foundation} &\textcolor{blue}{0.545} & 0.442 & 0.447 & 0.470 & PRIOR\cite{cheng2023prior} & \textcolor{blue}{34.15} & \textcolor{blue}{29.83} & 34.38 & MedBind$_{NM}$\cite{gao2024medbind} & 50.2 & 93.9 & 1.9 & 18.2\\
FLAIR-$\pi_{\text{EK}}$\cite{silva2025foundation} & 0.495 & \textcolor{red}{0.587} & 0.420 & \textcolor{blue}{0.667} & MGCA\cite{wang2022multi} & 33.45 & 29.38 & 34.01& MedBind$_{BD}$\cite{gao2024medbind} & \textcolor{red}{53.6} & 94.5 & 1.6 & \textcolor{red}{19.2} \\
\rowcolor[rgb]{0.925,0.957,1} M\textsuperscript{3}Bind & \textcolor{red}{0.564} & \textcolor{blue}{0.543} & \textcolor{blue}{0.480} & \textcolor{red}{0.672} & M\textsuperscript{3}Bind & \textcolor{red}{35.27} & \textcolor{red}{30.71} & \textcolor{blue}{34.96} & M\textsuperscript{3}Bind & \textcolor{blue}{52.4} & \textcolor{red}{95.6} & \textcolor{red}{2.3} & \textcolor{blue}{18.3}\\

\shline
\end{tabular}
\label{mesidor}
\end{large}%
}

\vspace{-2pt} 

\resizebox{\linewidth}{!}{%
\begin{small}
\begin{tabular}{l|cccc|c|cccccccc}
\shline
\multirow{3}{*}{\textbf{Methods}} & \multicolumn{4}{c|}{\textbf{Pathology}} & \multirow{3}{*}{\textbf{Methods}} & \multicolumn{8}{c}{\textbf{CT}} \\ \cline{2-5} \cline{7-14} 
 & \multicolumn{1}{c|}{\textbf{SkinCancer}\cite{kriegsmann2022deep}} & \multicolumn{1}{c|}{\textbf{Camelyon }\cite{veeling2018rotation}} & \multicolumn{1}{c|}{\textbf{SICAPv2 }\cite{silva2020going}} & \multicolumn{1}{c|}{\textbf{SkinTumor}\cite{kriegsmann2022deep}}  & & \multicolumn{4}{c|}{\textbf{CT-RATE}\cite{hamamci2024foundation}} & \multicolumn{4}{c}{\textbf{RAD-ChestCT}\cite{draelos2021machine}} \\ \cline{2-5} \cline{7-14} 
 & \multicolumn{4}{c|}{\textbf{Accuracy}} & & \textbf{AUC} & \textbf{F1} & \textbf{Acc} & \multicolumn{1}{c|}{\textbf{Prec}} & \textbf{AUC} & \textbf{F1} & \textbf{Acc} & \textbf{Prec} \\

\shline
CLIP\cite{radford2021learning} & 0.05 & 0.62 & 0.39 & 0.10 & CT-Net\cite{draelos2021machine} & 0.603 & 0.631 & 0.581 & 0.239 & 0.544 & 0.587 & 0.540&0.285 \\
BiomedCLIP\cite{zhang2023biomedclip} & 0.25 & 0.53 & \textcolor{red}{0.46} & 0.37 & CT-CLIP(ZS)\cite{hamamci2024foundation} & 0.731 & 0.707 & 0.668 & 0.323 & 0.629 & 0.642 & 0.595&0.336 \\
PLIP\cite{huang2023visual} & 0.37 & 0.59 & \textcolor{blue}{0.45} & 0.56 & CT-CLIP(VF)\cite{hamamci2024foundation} & \textcolor{red}{0.756} & \textcolor{blue}{0.738} & \textcolor{blue}{0.705} & \textcolor{red}{0.353} & \textcolor{red}{0.650} & \textcolor{blue}{0.659} & \textcolor{red}{0.659}&\textcolor{blue}{0.349} \\
QuiltNet\cite{ikezogwo2023quilt1m} & \textcolor{red}{0.45} & \textcolor{blue}{0.65} & 0.39 & \textcolor{blue}{0.58} & CT-CLIP(CF)\cite{hamamci2024foundation} & \textcolor{red}{0.756} & 0.724 & 0.689 & 0.339 & 0.643 & 0.649 & 0.607&0.340 \\
\rowcolor[rgb]{0.925,0.957,1} M\textsuperscript{3}Bind & \textcolor{red}{0.45} & \textcolor{red}{0.66} & 0.40 & \textcolor{red}{0.59} & M\textsuperscript{3}Bind & \textcolor{blue}{0.743} & \textcolor{red}{0.742} & \textcolor{red}{0.711} & \textcolor{blue}{0.348} & \textcolor{blue}{0648} & \textcolor{red}{0.662} & \textcolor{blue}{0.621}&\textcolor{red}{0.352} \\

\shline
\end{tabular}
\end{small}%
}
\label{Pathology}
\end{table*}

\section{Experiments}
\subsection{Pretrained CLIP-like Models Selected}

In our experiments, we select the following pre-trained CLIP-like models to train our M\textsuperscript{3}Bind, namely MGCA \cite{wang2022multi} (specific to the X-Ray image), QuiltNet \cite{ikezogwo2023quilt1m} (specific to the pathology image), FLAIR \cite{silva2025foundation} (specific to the retina image), and CT-CLIP \cite{hamamci2024foundation} (specific to CT volumes). 
Regarding the ECG modality, since currently no related CLIP-like work is available, for subsequent fair comparisons, we follow the definition of the MedBind model \cite{gao2024medbind} and train an ECG-CLIP model ourselves, which is then used for M\textsuperscript{3}Bind training.
We will primarily compare our approach with the selected CLIP-like models to demonstrate that our pre-training framework not only enables implicit alignment across different image modalities for these models, but also enhances their original performance.

\subsection{Experimental Settings}
For all experiments, we maintained consistent preprocessing protocols as specified in the original papers for each modality-specific model (FLAIR, MGCA, CT-CLIP, ECG-CLIP, and QuiltNet) \cite{silva2025foundation, wang2022multi, hamamci2024foundation, ikezogwo2023quilt1m}. 
For the pre-training phase, we employed the AdamW optimizer with a base learning rate of 2e-5, incorporating warm-up and cosine decay schedules to ensure stable convergence. 
We configured batch sizes according to the complexity of the model and hardware constraints, namely 72 for most, with the exception of CT-CLIP \cite{hamamci2024foundation} which required a reduced batch size of 8 due to its larger memory footprint. 
Text-modality alignment was performed with a batch size of 64 in all domains, using our adaptive sampling strategy to balance representation in datasets of different sizes. 
Training was carried out for 15,000 iterations with the balance factor $\lambda$ set to 10, prioritizing inter-modality text alignment before connecting the individual medical imaging modalities through the unified text space.
In the SESKD training, we utilized a pretrained BioClinicalBERT model \cite{alsentzer2019publicly} as our student text encoder. 
The training process was divided into two stages: first, we applied the knowledge distillation loss $\mathcal{L}_\text{KD}$ for 1,200 iterations to establish baseline alignment, then integrated the complete SESKD objective ($\mathcal{L}_\text{SESKD}$) for an additional 1,200 iterations to preserve both embedding similarity and contrastive relationships. 
All other optimization parameters remained consistent with the previous.
All experiments were conducted on seven RTX 4090 GPUs, while the SESKD phase requires an additional RTX 4090 GPU.

\subsection{Evaluations in Zero-Shot Manner}

\noindent \textbf{For the Retina modality:}
Experiments on retina images were performed following FLAIR's settings, which involved prompt mappings and disease-specific descriptions to enhance domain knowledge. Validation was performed on MESIDOR and ODIR datasets to test domain shift and novel class generalization. Comparisons were made with FLAIR-$\pi_{\text{naive}}$, FLAIR-$\pi_{\text{EK}}$, CLIP, and BiomedCLIP \cite{zhang2023biomedclip}. Our M\textsuperscript{3}Bind achieved ACA scores \cite{zhao2019bira} of 0.564 and 0.543 on MESIDOR ($\pi_{naive}$ and $\pi_{EK}$, respectively) and 0.480 and 0.672 on ODIR200x3, demonstrating superior performance in both metrics.

\noindent \textbf{For the X-Ray modality:} MIMIC-CXR pre-trained models were used to predict CheXpert labels, with performance evaluated on the CheXpert 5x200 test set using accuracy, F1 score, and precision metrics. The baseline comparisons shown in Table \ref{mesidor} included ConVIRT \cite{zhang2022contrastive}, GLoRIA, MGCA, and PRIOR\cite{cheng2023prior}. 
Our model achieved the highest scores, with an accuracy of 35.27, an F1-score of 30.71, and a precision of 34.96.

\noindent \textbf{For the ECG modality:} 
Combining the performance in the MIMIC-ECG internal validation dataset and the PTB-XL external validation dataset, we achieve the best performance. In contrast, MedBind$_\text{NM}$, which trains directly from randomly initialized networks, results in a performance loss. Even after introducing a loss function to align ECG and X-Ray, MedBind$_\text{BD}$'s performance advantages are not significant.

\noindent \textbf{For the Pathology modality:} In addition to CLIP, BiomedCLIP, and QuiltNet, we also compared our approach with PLIP \cite{huang2023visual}, another vision-language pre-training model for pathology. Apart from the SkinCancer dataset, we demonstrate performance improvements over QuiltNet on our other datasets, with overall performance exceeding any baseline.

\noindent \textbf{For the CT modality:} Internal validation was carried out on CT-RATE and external validation on RAD-ChestCT. Baseline comparisons included CT-Net\cite{draelos2021machine} and three variants of CT-CLIP (ZS, VF, CF). Our model achieved F1 scores of 0.742 in CLIP-RATE and 0.662 on RAD-ChestCT, reflecting high performance in both datasets.

\noindent Overall, M\textsuperscript{3}Bind either outperforms or comes close to outperforming in most cases. Even when it does not win, it does not show a noticeable performance loss compared to the pre-trained models we used.

\begin{table}[h]
\centering
\setlength{\abovecaptionskip}{0pt} 
\setlength{\belowcaptionskip}{0pt} 
\caption{Few-Shot classification performance (Accuracy \%) comparison of M\textsuperscript{3}Bind across different modalities in terms of accuracy (mean $\pm$ standard deviation) against modality-specific state-of-the-art models on ten datasets spanning five distinct medical imaging modalities.
Evaluations are conducted with limited training examples to assess cross-modal transfer capabilities.
Red and blue highlighting indicates the best and second-best performance, respectively.}
\label{tab:sys_compare}

\begin{minipage}{1\linewidth} 
    \resizebox{\linewidth}{!}{%
    \begin{small}
    \begin{tabular}{p{2cm}|cccccc}
    \shline
    \multirow{3}{*}{\textbf{Methods}} & \multicolumn{6}{c}{\textbf{Retina}} \\ \cline{2-7}
    & \multicolumn{3}{c|}{\textbf{MESIDOR}} & \multicolumn{3}{c}{\textbf{ODIR200x3}} \\ \cline{2-7} 
     & \textbf{1} & \textbf{5} & \textbf{10} & \textbf{1} & \textbf{5} & \textbf{10} \\
    
    \shline
CLIP & $0.299 \pm 0.02$ & $0.420 \pm 0.03$ & $0.425 \pm 0.04$ & $0.686 \pm 0.02$ & $0.810 \pm 0.03$ & $0.823 \pm 0.04$ \\
BiomedCLIP & $0.312 \pm 0.03$ & $0.430 \pm 0.04$ & $0.434 \pm 0.03$ & $0.712 \pm 0.02$ & $0.847 \pm 0.03$ & $0.846 \pm 0.04$  \\
ImageNet Init & $0.293 \pm 0.02$ & $0.413 \pm 0.03$ & $0.407 \pm 0.05$ & $0.615 \pm 0.02$ & $0.805 \pm 0.03$ & $0.813 \pm 0.04$ \\
FLAIR- $\pi_{\text{EK}}$ & \textcolor{red}{$0.483 \pm 0.05$} & \textcolor{blue}{$0.601 \pm 0.04$} & \textcolor{blue}{$0.605 \pm 0.03$} & \textcolor{blue}{$0.784 \pm 0.03$} & \textcolor{blue}{$0.877 \pm 0.04$} & \textcolor{blue}{$0.875 \pm 0.05$} \\
\rowcolor[rgb]{0.925,0.957,1} OURS & \textcolor{blue}{$0.467 \pm 0.04$} & \textcolor{red}{$0.613 \pm 0.03$} & \textcolor{red}{$0.642 \pm 0.05$} & \textcolor{red}{$0.802 \pm 0.03$} & \textcolor{red}{$0.890 \pm 0.04$} & \textcolor{red}{$0.903 \pm 0.05$} \\

    \shline
    \end{tabular}
    \end{small}%
    }

    \vspace{-2pt} 

    \resizebox{\linewidth}{!}{%
    \begin{small}
    \begin{tabular}{p{2cm}|cccccc}
    \shline
    \multirow{3}{*}{\textbf{Methods}} & \multicolumn{6}{c}{\textbf{X-Ray}} \\ \cline{2-7}
    & \multicolumn{3}{c|}{\textbf{RSNA}} & \multicolumn{3}{c}{\textbf{CheXpert}} \\ \cline{2-7} 
     & \textbf{1} & \textbf{10} & \textbf{100} & \textbf{1} & \textbf{10} & \textbf{100} \\
    
    \shline
ConVIRT & $84.17 \pm 0.77$ & $86.92 \pm 0.13$ & $88.74 \pm 0.36$ & $85.02 \pm 0.28$ & $87.58 \pm 0.53$ & $88.21 \pm 0.46$ \\
GLoRIA  & $84.12 \pm 0.47$ & $86.83 \pm 0.53$ & $89.13 \pm 0.12$ & $83.61 \pm 0.52$ & $87.40 \pm 0.39$ & $88.34 \pm 0.12$  \\
PRIOR   & \textcolor{blue}{$85.74 \pm 0.36$} & $87.08 \pm 0.19$ & $89.22 \pm 0.16$ & \textcolor{blue}{$86.16 \pm 0.64$} & \textcolor{blue}{$88.31 \pm 0.20$} & \textcolor{blue}{$88.61 \pm 0.29$} \\
MGCA    & \textcolor{red}{$85.80 \pm 0.68$} & \textcolor{red}{$87.66 \pm 0.21$} & \textcolor{blue}{$89.30 \pm 0.16$} & $85.63 \pm 0.33$ & $87.65 \pm 0.33$ & $88.30 \pm 1.48$ \\
\rowcolor[rgb]{0.925,0.957,1} OURS    & $84.48 \pm 0.12$ & \textcolor{blue}{$87.54 \pm 0.24$} & \textcolor{red}{$89.35 \pm 0.43$} & \textcolor{red}{$88.07 \pm 0.52$} & \textcolor{red}{$88.64 \pm 0.27$} & \textcolor{red}{$88.98 \pm 0.39$} \\

    \shline
    \end{tabular}
    \end{small}%
    }

    \vspace{-2pt} 

    \resizebox{\linewidth}{!}{%
    \begin{small}
    \begin{tabular}{p{2cm}|cccccc}
    \shline
    \multirow{3}{*}{\textbf{Methods}} & \multicolumn{6}{c}{\textbf{ECG}} \\ \cline{2-7}
    & \multicolumn{3}{c|}{\textbf{PTB-XL}\cite{wagner2020ptb}} & \multicolumn{3}{c}{\textbf{ICBEB}\cite{liu2018:icbeb}} \\ \cline{2-7} 
     & \textbf{1} & \textbf{4} & \textbf{8} & \textbf{1} & \textbf{4} & \textbf{8} \\
    
    \shline

ECG-CLIP  & $38.83 \pm 1.42$ & $50.17 \pm 0.73$ & \textcolor{blue}{$54.37 \pm 0.80$} & \textcolor{blue}{$19.72 \pm 0.26$} & $25.49 \pm 0.48$ & $27.44 \pm 0.27$  \\
MedBind$_{NM}$\cite{gao2024medbind}   & \textcolor{blue}{$39.03 \pm -.-$} & $49.33 \pm -.-$ & $54.13
\pm -.-$ & $19.66 \pm -.-$ & $25.31 \pm -.-$ & $27.38 \pm -.-$ \\
MedBind$_{BD}$\cite{gao2024medbind}    & \textcolor{red}{$39.80 \pm -.-$} & \textcolor{blue}{$50.40 \pm -.-$} & $53.43 \pm -.-$ & $19.97 \pm -.-$ & \textcolor{red}{$26.04 \pm -.-$} & \textcolor{red}{$28.08 \pm -.-$} \\
\rowcolor[rgb]{0.925,0.957,1} M\textsuperscript{3}Bind    & $39.48 \pm 0.99$ & \textcolor{red}{$50.55 \pm 0.31$} & \textcolor{red}{$54.50 \pm 0.64$} & \textcolor{red}{$20.03 \pm 1.86$} & \textcolor{blue}{$25.84 \pm 0.41$} & \textcolor{blue}{$27.82 \pm 0.17$} \\

    \shline
    \end{tabular}
    \end{small}%
    }

      \vspace{-2pt} 

    \resizebox{\linewidth}{!}{%
    \begin{small}
    \begin{tabular}{p{2cm}|cccccc}
    \shline
    \multirow{3}{*}{\textbf{Methods}} & \multicolumn{6}{c}{\textbf{Pathology}} \\ \cline{2-7}
    & \multicolumn{3}{c|}{\textbf{Camelyon}} & \multicolumn{3}{c}{\textbf{SICAPv2}} \\ \cline{2-7} 
     & \textbf{1} & \textbf{10} & \textbf{100} & \textbf{1} & \textbf{10} & \textbf{100} \\
    
    \shline
CLIP       & $80.73 \pm 0.53$                    & $81.42 \pm 0.21$                  & $82.07 \pm 0.14$    
           & $52.64 \pm 1.91$                    & $62.79 \pm 0.57$ & $65.94 \pm 0.43$ \\
PLIP       & $87.26 \pm 0.61$  & $87.42 \pm 0.23$  & $87.45 \pm 0.19$ & $65.49 \pm 0.95$ & $68.97 \pm 0.63$ & $73.15 \pm 0.71$  \\
PLIP(FT)   & $87.50 \pm 0.35$                   & $87.57 \pm 0.18$                   & $87.53 \pm 0.07$ 
           & \textcolor{blue}{$69.54 \pm 0.47$}                   & $73.18 \pm 0.61$                   & $74.85 \pm 0.53$ \\
QuiltNet   & \textcolor{blue}{$87.51 \pm 0.32$}                   & \textcolor{blue}{$87.40 \pm 0.13$}                   & \textcolor{blue}{$87.27 \pm 0.09$} 
           & \textcolor{red}{$69.73 \pm 0.78$}                   & \textcolor{blue}{$74.19 \pm 0.48$}                   & \textcolor{blue}{$75.31 \pm 0.36$} \\
OURS       & \textcolor{red}{$87.65 \pm 0.41$} & \textcolor{red}{$87.88 \pm 0.37$} & \textcolor{red}{$87.63 \pm 0.06$} & $68.91 \pm 1.12$ & \textcolor{red}{$74.56 \pm 0.68$} & \textcolor{red}{$75.54 \pm 0.50$} \\

    \shline
    \end{tabular}
    \end{small}%
    }

      \vspace{-2pt} 

    \resizebox{\linewidth}{!}{%
    \begin{small}
    \begin{tabular}{p{2cm}|cccccc}
    \shline
    \multirow{3}{*}{\textbf{Methods}} & \multicolumn{6}{c}{\textbf{CT}} \\ \cline{2-7}
    & \multicolumn{3}{c|}{\textbf{CT-RATE}} & \multicolumn{3}{c}{\textbf{RAD-ChestCT}} \\ \cline{2-7} 
     & \textbf{1} & \textbf{5} & \textbf{10} & \textbf{1} & \textbf{5} & \textbf{10} \\
    
    \shline
CT-Net     & $63.12 \pm 0.53$ & $67.56 \pm 0.61$ & $69.75 \pm 0.47$ & $58.79 \pm 0.73$ & $63.86 \pm 0.55$ & $67.65 \pm 0.63$ \\
CT-CLIP(ZS) & $75.85 \pm 0.68$ & $81.17 \pm 0.79$ & $82.42 \pm 0.57$ & $65.37 \pm 0.81$ & $71.57 \pm 0.42$ & $74.58 \pm 0.59$  \\
CT-CLIP(VF) & \textcolor{blue}{$78.34 \pm 0.49$} & \textcolor{blue}{$82.99 \pm 0.51$} & \textcolor{red}{$83.66 \pm 0.73$} & \textcolor{blue}{$68.02 \pm 0.67$} & \textcolor{blue}{$72.45 \pm 0.72$} & \textcolor{blue}{$75.42 \pm 0.53$} \\
CT-CLIP(CF) & $78.08 \pm 0.65$ & $82.85 \pm 0.58$ & \textcolor{blue}{$83.58 \pm 0.79$} & $67.81 \pm 0.51$ & $72.28 \pm 0.43$ & $75.13 \pm 0.77$ \\
\rowcolor[rgb]{0.925,0.957,1} OURS       & \textcolor{red}{$78.77 \pm 0.52$} & \textcolor{red}{$83.20 \pm 0.46$} & $83.39 \pm 0.61$ & \textcolor{red}{$68.08 \pm 0.41$} & \textcolor{red}{$72.94 \pm 0.69$} & \textcolor{red}{$75.82 \pm 0.57$} \\

    \shline
    \end{tabular}
    \end{small}%
    }
\end{minipage}
\label{Few shot}
\end{table}

\subsection{Evaluations in Few-Shot Manner}

In this evaluation, for better comparison, we used different few-shot ratios in various modalities, including 1, 5, 10 for the retina and CT images, 1, 4, 8 for ECG, and 1, 10, 100 for X-ray and pathology images. 
Our M\textsuperscript{3}Bind demonstrated strong cross-modal adaptability, achieving state-of-the-art performance in most benchmarks, as shown in Table \ref{Few shot}. 
For retinal imaging, M\textsuperscript{3}Bind achieved superior performance in both MESIDOR (61.3\% at 5-shot, 62.0\% at 10-shot) and ODIR200x3 (80.2\% at 1-shot, 89.2\% at 10-shot), outperforming BiomedCLIP and FLAIR-$\pi_{EK}$. 
In X-Ray analysis, we attained the highest accuracy on CheXpert across all shots (88.07\%/ 88.64\%/ 88.98\%), surpassing prior specialized models like PRIOR and MGCA. 
Although the original PRIOR paper mentions that MGCA neglects sentence-level clinical information, which is important for complex downstream tasks like multi-label classification on CheXpert, leading to performance that is not as good as expected, we not only surpass the pre-trained model MGCA we use but also directly exceed PRIOR. This gap is particularly evident in the 1-shot scenario.
The model showed particular strength in CT interpretation, achieving 78.77\% (1-shot) and 83.20\% (5-shot) on CT-RATE, and 75.82\% (10-shot) on RAD-ChestCT, outperforming all CT-CLIP variants. 
Notably, our approach demonstrated exceptional ECG classification capabilities on PTB-XL (50.55\% at 4-shot, 54.50\% at 8-shot) and ICBEB (20.03\% at 1-shot), exceeding MedBind variants. 
In pathology images, M\textsuperscript{3}Bind achieved record performance on Camelyon (87.65-87.88\%) and competitive results on SICAPv2 (74.56\% at 10-shot), highlighting cross-modal robustness. 
Our method maintained strong scaling capabilities, with performance improvements over baselines in extreme low-shot scenarios (1-10 samples), while showing diminishing returns beyond 100 shots, suggesting effective few-shot knowledge transfer.


\begin{figure*}[t]
    \centering
    \includegraphics[width=0.9\textwidth]{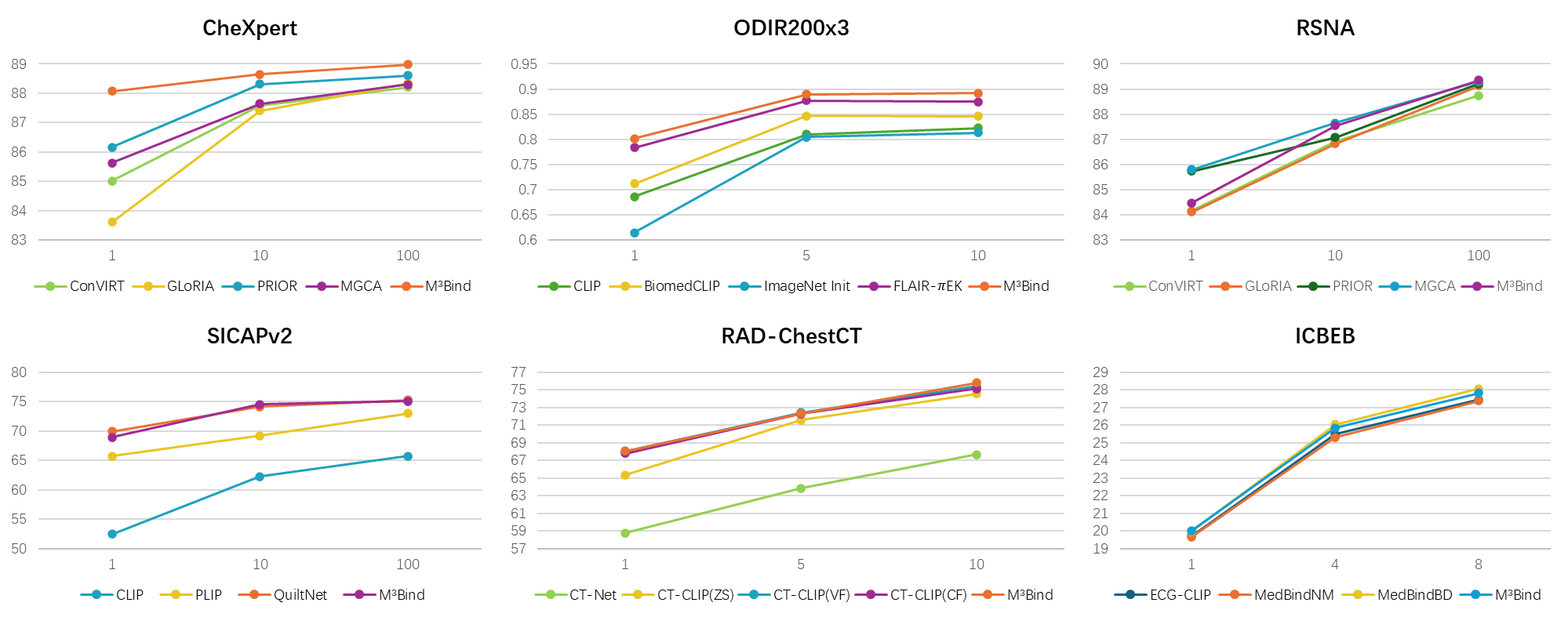} 
    \caption{Comparison of Few-Shot Performance Across Various Datasets (ODIR200x3, CheXpert, SICAPv2, RAD-ChestCT) With Increasing Shot Numbers.}
    \label{fig:algorithm_Few}
\end{figure*}

\noindent Figure \ref{fig:algorithm_Few} illustrates the exceptional few-shot learning efficacy of our methodology across several medical imaging modalities, demonstrating superior performance in handling diverse datasets, including ODIR200x3, CheXpert, SICAPv2, and RAD-ChestCT.

\subsection{Evaluations in Cross-Modal Manner}

\begin{table}[h]
\centering
\caption{
Cross-modal retrieval performance (Recall Top@1, 5, 10) comparison of M\textsuperscript{3}Bind against two different variants of MedBind. Red and blue highlighting indicate the best and second-best performance, respectively.}
\label{cross}
\resizebox{0.75\linewidth}{!}{
\begin{tabular}{c|ccc}
\toprule
\multirow{2}{*}{\bfseries Methods} & \multicolumn{3}{c}{\bfseries MIMIC-PAIR} \\ \cline{2-4} 
 & Top@1 & Top@5 & Top@10 \\
\midrule
 MedBind$_{NM}$\cite{gao2024medbind} & 1.36 & 7.86 & 15.00 \\
 MedBind$_{BD}$\cite{gao2024medbind} & \textcolor{red}{12.86} & \textcolor{red}{30.95} & \textcolor{red}{55.43} \\
 M\textsuperscript{3}Bind  & \textcolor{blue}{1.82} & \textcolor{blue}{9.64} & \textcolor{blue}{17.29} \\
\bottomrule
\end{tabular}
}
\end{table}

To further evaluate the performance of cross-modal alignment between different image modalities, we conducted a cross-modal retrieval task from X-Ray to ECG modalities on the MIMIC-PAIR dataset. 
According to the experimental results in Table~\ref{cross}, our framework outperforms MedBind$_{NM}$ in Recall Top@1, Recall Top@5, and Recall Top@10, validating that introducing more modalities and more data indeed enhances cross-modal performance.

Notably, the results for MedBind$_{BD}$ are significantly better than both our method and MedBind$_{NM}$. This is because MedBind$_{BD}$ uses the MIMIC-PAIR training set during the pre-training phase and introduces a loss function to align the X-ray and ECG modalities. Therefore, MedBind$_{BD}$ does not perform this task in a zero-shot manner, making this comparison unfair, and these results should not be considered. We have also discussed the approach of MedBind$_{BD}$, noting that as the number of modalities increases, the dataset and computational resource requirements for pairwise alignment between modalities become impractical.

\subsection{Ablation Study}
\noindent \textbf{Ablation Analysis of X-Ray, ECG, Retina, Pathology, and CT Modalities.} We conducted an ablation study to evaluate the contributions of X-ray, ECG, retina, pathology and CT images on the ODIR200x3\footref{1} and MIMIC-PAIR datasets, as summarized in Table~\ref{t3}. 
As the number of modalities involved in training increases, the model's performance gradually improves, especially when the CT modality, which is highly relevant to chest X-Ray and ECG, is added. The experimental results demonstrate that each additional modality incrementally improves the quality of the shared space representation.

\noindent \textbf{Ablation Study of Adaptive Modality Balancing and Shared Embedding Space Knowledge Distillation Modules.} We conducted an ablation study to investigate the individual contributions of the Adaptive Modality Balancing (AMB) and Shared Embedding Space Knowledge Distillation (SESKD) modules within the M$^3$Bind framework, with performance results shown in Table~\ref{t4}. 
The results indicate that, in the image-to-text zero-shot classification task, the AMB module can effectively improve model performance, while the effect of SESKD is not significant and may even slightly decrease performance. However, in the cross-modal retrieval task between X-Ray and ECG, both modules are very effective. This shows that the AMB can prevent biased alignment, while SESKD can unify the representation of the shared space.

 \begin{table}[h]
\centering
\caption{Ablation Study on Individual Contributions of X-Ray, ECG, Retina, Pathology, and CT Modalities: Performance is evaluated on the MIMIC-PAIR datasets.}
\label{t3}
\resizebox{\linewidth}{!}{
\begin{tabular}{ccccc|ccc}
\toprule
\multirow{2}{*}{\bfseries X-Ray} & \multirow{2}{*}{\bfseries ECG} & \multirow{2}{*}{\bfseries Retina} & \multirow{2}{*}{\bfseries Pathology} & \multirow{2}{*}{\bfseries CT} & \multicolumn{3}{c}{\bfseries MIMIC-PAIR} \\ \cline{6-8} 
 &  &  &  &  & Top@1 & Top@5 & Top@10 \\
\midrule

 \checkmark & \checkmark &  & & & 1.25 & 7.03 & 15.14 \\

 \checkmark & \checkmark & \checkmark &  & & 1.19 & 7.62 & 14.99 \\

 \checkmark & \checkmark & \checkmark & \checkmark & & 1.48 & 8.21  & 15.73   \\
 
 \checkmark & \checkmark & \checkmark & \checkmark & \checkmark &  \textcolor{red}{1.82} & \textcolor{red}{9.64} & \textcolor{red}{17.29} \\
\bottomrule
\end{tabular}
}
\end{table}

\begin{table}[h]
\centering
\caption{Ablation Study of AMB and SESKD Modules: Performance Evaluation on the ODIR200x3 and MIMIC-PAIR Datasets Using the M$^3$Bind Framework.}
\label{t4}
\resizebox{\linewidth}{!}{
\begin{tabular}{ccc|cc|ccc}
\toprule
\multirow{2}{*}{\bfseries M$^3$Bind} & \multirow{2}{*}{\bfseries AMB} & \multirow{2}{*}{\bfseries SESKD} & \multicolumn{2}{c|}{\bfseries ODIR200x3\footref{1}} & \multicolumn{3}{c}{\bfseries MIMIC-PAIR} \\ \cline{4-5} \cline{6-8}
 &  &  & $\pi_{\text{naive}}$ & $\pi_{\text{EK}}$ & Top@1 & Top@5 & Top@10 \\
\midrule
\checkmark & & & 0.452 & 0.622 & 1.29 & 6.85 & 15.52 \\
\checkmark & \checkmark & & \textcolor{red}{0.478} & 0.659 & 1.65 & 8.92 & 17.01 \\
\checkmark & & \checkmark & 0.440 & 0.617 & 1.40 & 8.27 & 15.88 \\
\checkmark & \checkmark & \checkmark & 0.472 & \textcolor{red}{0.667} & \textcolor{red}{1.82} & \textcolor{red}{9.64} & \textcolor{red}{17.29} \\
\bottomrule
\end{tabular}
}
\end{table}
\section{Conclusion}
In this paper, we introduce M\textsuperscript{3}Bind, a novel multimodal pre-training method for aligning medical imaging representation. 
Using an anchored text modality, M\textsuperscript{3}Bind harmonizes image representations across five CLIP-like medical models, allowing cross-modal alignment without relying on arbitrary paired data between two medical image modalities. 
Experimental results show that our framework not only achieves implicit alignment across different image modalities, but also brings performance improvements to the original models on image-text tasks. The flexibility of our framework allows it to benefit from stronger CLIP-like models—if we select more powerful base models, our framework can accordingly achieve even better performance, potentially reaching state-of-the-art (SOTA) results.
%
%
%
%
{
    \small
    \bibliographystyle{ieeenat_fullname}
    \bibliography{main}
}

\end{document}